\documentclass[12pt,epsf]{iopart}
\usepackage{graphicx}
\usepackage{amssymb,latexsym}
\usepackage{amsfonts}
\usepackage{amssymb}
\usepackage{cancel}
\usepackage{color}
\input{colordvi.tex}

\newcommand{\bear}{\begin{array}}  \newcommand{\eear}{\end{array}}
\newcommand{\bea}{\begin{eqnarray}}  \newcommand{\eea}{\end{eqnarray}}
\newcommand{\beq}{\begin{equation}}  \newcommand{\eeq}{\end{equation}}
\newcommand{\bef}{\begin{figure}}  \newcommand{\eef}{\end{figure}}
\newcommand{\bec}{\begin{center}}  \newcommand{\eec}{\end{center}}
  
\newcommand{\lmk}{\left(}  \newcommand{\rmk}{\right)}

\newcommand{\del}{\partial}

\newcommand{\beqa}{\begin{eqnarray}}
\newcommand{\eeqa}{\end{eqnarray}}
\newcommand{\p}{\phi}

\newcommand{\ka}{\kappa}

\newcommand{\Om}{\Omega}
\newcommand{\hatg}{\widehat{g}}
\newcommand{\hatV}{\widehat{V}}
\newcommand{\hatp}{\widehat{\phi}}
\newcommand{\hatep}{\widehat{\epsilon}}
\newcommand{\hateta}{\widehat{\eta}}

\newcommand{\hatH}{\widehat{H}}
\newcommand{\hath}{\widehat{h}}

\newcommand{\hatN}{\widehat{N}}

\newcommand{\hatnabla}{\widehat{\nabla}}
\newcommand{\hatpsi}{\widehat{\psi}}
\newcommand{\R}{{\cal R}}

\begin{document}

\title{{Extended Slow-Roll Conditions and Primordial Fluctuations:
Multiple Scalar Fields and Generalized Gravity}}

\author{Takeshi Chiba}%
\address{
Department of Physics, \\
College of Humanities and Sciences, \\
Nihon University, \\
Tokyo 156-8550, Japan}
\author{Masahide Yamaguchi}%
\address{Department of Physics and Mathematics, Aoyama Gakuin
University, Sagamihara 229-8558, Japan \\ and \\
Department of Physics, Stanford University, Stanford CA 94305}

\date{\today}


\begin{abstract}
As an extension of our previous study, we derive slow-roll conditions
for multiple scalar fields which are non-minimally coupled with gravity
and for generalized gravity theories of the form $f(\phi,R)$. We provide
simple formulae of the spectral indices of scalar/tensor perturbations
in terms of the slow-roll parameters. 
\end{abstract}

\maketitle

\section{Introduction}

Inflation provides not only for a compelling explanation for the
homogeneity and isotropy of the universe but also for the observed
spectrum of density perturbations \cite{linde,ll,lr}.  In its simplest
realization, slow-roll inflation predicts an almost scale invariant
spectrum of density perturbations and an almost scale invariant spectrum
of gravitational waves.  Since the observed scalar spectral index is
already close to unity \cite{wmap}, the slow-roll approximation is
useful approximation to the reality.  Then, in confrontation with
observational data, it is useful to widen the range of models of
slow-roll inflation to connect the observational data with microphysics
models of inflation.

In our previous paper \cite{cy}, we have derived the slow-roll
conditions for a single scalar field which non-minimally couples to
gravity and provided the formulae of the scalar/tensor spectral indices
in terms of the slow-roll parameters. In this paper, extending our
previous study, we derive the slow-roll conditions for non-minimally
coupled multiple scalar fields and provide simple formulae of the
spectral indices in terms of the slow-roll parameters. This is natural
extension in light of microphysics models of inflation because the
existence of many light scalars is predicted in high energy theories
like supergravity or superstring. We also derive the slow-roll
conditions for generalized gravity models and give the formulae of
spectral indices.

In Sec.2, we derive the slow-roll conditions for multiple-field models
of inflation.  We give simple formulae of spectral indices and compare
them with those in the Einstein frame.  We give an extended assisted
inflation model as an example. In Sec.3, we consider generalized
gravity models of the form $f(\p,R)$ and find that to the first order in
the slow-roll approximations, the system is reduced to a single field
with non-minimal coupling, to which our previous results \cite{cy} can
be applied. In Appendix, we give several formulae for multiple scalar fields and 
for generalized gravity models in the Einstein frame
which are useful for calculations in the text.

\section{Slow-Roll Inflation with a Non-minimally Coupled Multiple Scalar Fields}

We consider the multi-field model of inflation which couples
non-minimally to gravity. The action in the Jordan frame metric
$g_{\mu\nu}$ is
\beq
S=\int d^4x\sqrt{-g}\left[{\Omega(\phi)\over 2\ka^2}R-
{1\over 2}h_{ab}g^{\mu\nu}\del_{\mu}\phi^a \del_{\nu}\phi^b-V(\phi)\right],
\label{action}
\eeq
where greek indices $\mu,\nu,\dots =0,1,2,3$ denote spacetime indices;
latin indices $a,b,c, \dots=1,2,\dots ,n$ serve to label the $n$ scalar
fields and $h_{ab}$ is a metric on the scalar field space. $\ka^2\equiv
8\pi G$ is the bare gravitational constant and $\Omega(\p)R$ term
corresponds to the non-minimal coupling of the scalar fields to
gravity. We assume that the scalar field metric $h_{ab}$ is  not degenerate. 
An interesting exception is a generalized gravity of the form $f(\phi,R)$
the case which will be considered in Sec.3.

We assume that the universe is described by the flat, homogeneous, and
isotropic universe model with the scale factor $a$. The field equations
are then given by
\beqa
H^2\Om+H\dot\Om=\frac{\ka^2}{3}\left(\frac{1}{2}\dot\p^a \dot\p_a+V\right),\\
\ddot\Om-H\dot\Om+2\dot H\Om=-\ka^2\dot\p^a \dot\p_a,\label{eq:2}\\
\frac{D}{dt}\dot\p^a+3H\dot\p^a+V^{,a}
   -\frac{3\Om^{,a}}{\ka^2}(\dot H+2H^2)=0,
\eeqa
where the dot denotes the derivative with respect to the cosmic time,
$(D/dt)\dot\phi^a=\dot\phi^b\nabla_b\dot\phi^a=\ddot\phi^a+\Gamma^{a}_{bc}\dot\phi^b\dot\phi^c$,
$\nabla_a$ is the covariant derivative associated with $h_{ab}$ and
$V^{,a}= h^{ab}\del V/\del{\phi^b}$.

\subsection{Slow-Roll Conditions}

Under the slow-roll approximations, the time scale of the motion of the
scalar fields is assumed to be much larger than the cosmic time scale
$H^{-1}$.  As an extended slow-rolling of the scalar fields, we assume
that $\dot\p^a\dot\p_a \ll V, |\dot\Om|\ll H\Om$, 
$|(D/dt)\dot\p^a|\ll H|\dot\p^a|$, $|(D/dt)\dot\p^a|\ll
|V^{,a}|$, then we obtain
\beqa
&&H^2\Om\simeq \frac{\ka^2}{3}V,
\label{eq:slow:1}\\
&&3H \dot\p^a\simeq -V^{,a}+6\frac{\Om^{,a}}{\ka^2}H^2\simeq -
\Om^2\left(\frac{V}{\Om^2}\right)^{,a}:=-V_{eff}^{,a},
\label{eq:slow:2}
\eeqa
where in the second equation, we have assumed $|\dot H/H^2|\ll 1$ which
should be checked later.

In the following, we derive the consistency conditions for the extended
slow-rolling of the scalar field (the extended slow-roll conditions). By
computing $(D/dt)\dot\p$ from Eq.~(\ref{eq:slow:2}), we obtain
\beqa
\frac{(D/dt)\dot\p^a}{H\dot\p^a}\simeq 
  -\frac{\dot H}{H^2}-\frac{\dot\p^b\nabla_b\nabla^aV_{eff}}{3H^2\dot\p^a}
  \simeq -\frac{\dot H}{H^2}
  - \frac{\Om \nabla_b\nabla^aV_{eff}}{\kappa^2 V}\frac{V_{eff}^{,b}}{V_{eff}^{,a}}.
\label{ratio:1}
\eeqa
Also, we note
\beqa
&& \frac{(D/dt)\dot\p^a}{V^{,a}}
  \simeq -\frac{(D/dt)\dot\p^a}{3H\dot\p^b}\frac{V_{eff}^{,b}}{V^{,a}}.
\eeqa
Moreover, from Eqs.~(\ref{eq:slow:1}) and (\ref{eq:slow:2}),
\beqa
&&\frac{\dot\p^a\dot\p_a}{V}\simeq 
  \frac{\Om V_{eff}^{,a} V_{eff,a}}{3\ka^2 V^2},\label{ratio:2}\\
&&\frac{\dot\Om}{H\Om}\simeq -\frac{\Om_{,a}V_{eff}^{,a}}{\ka^2 V}.\label{ratio:3}
\eeqa  
Hence, we finally introduce the following three slow-roll parameters and
obtain the extended slow-roll conditions:
\beqa
&&\epsilon:=\frac{\Om V_{eff,a}V_{eff}^{,a}}{2\ka^2 V^2};~~~~\epsilon \ll 1,
\label{slow-roll:1}\\
&&\eta_{ab}:=\frac{\Om \nabla_a\nabla_bV_{eff}}{\ka^2 V};~~~~|{\eta_{a}}^{b}|\ll1,
\label{slow-roll:2}\\
&&\delta:=\frac{\Om_{,a}V_{eff}^{,a}}{\ka^2 V};~~~~|\delta|\ll1,
\label{slow-roll:3}
\eeqa
and one subsidiary condition:
\beqa
\left|\frac{V_{eff}^{,a}}{V^{,b}}\right|= {\cal O}(1).\label{sub:1}
\eeqa
Note that since from Eq.~(\ref{eq:slow:1}) $\dot H/H^2$ is approximated as
\beqa
 \frac{\dot H}{H^2} \simeq 
- \frac{3\dot\p^a\dot\p_a}{2V}+ \frac{\dot \Omega}{2H\Omega} 
=-\epsilon-{1\over 2}\delta.
\label{hdot}
\eeqa
Thus, $|\dot H/H^2|\ll 1$ is guaranteed by these slow-roll conditions.

To sum up, the extended slow-roll conditions consist of three main
conditions Eqs.~(\ref{slow-roll:1}), (\ref{slow-roll:2}),
(\ref{slow-roll:3}) and one subsidiary condition Eq.~(\ref{sub:1}).\footnote{It is 
to be noted that the subsidiary
condition is the sufficient condition for slow-roll and the necessary
condition is that Eq.~(\ref{sub:1}) multiplied by $\epsilon,
\eta_{ab}$ or $\delta$ is sufficiently small.}

\subsection{Perturbations}

In \cite{cy,ms}, it is shown that the gauge invariant curvature
perturbation $\R$ is invariant under the conformal transformation into
the Einstein frame. Then the power spectrum $P_{\R}(k)$ in the Jordan
frame is given by that in the Einstein frame \cite{alex,ss,ns},
\beqa
P_{\R}(k)=\left(\frac{\hatH}{2\pi}\right)^2\hath^{ab}
\hatN_{,a}\hatN_{,b}
=\left(\frac{H}{2\pi}\right)^2N_{,a}N^{,a},
\label{spec:scalar}
\eeqa
where $k$ is a comoving wavenumber at the horizon exit ($k=aH$) and the
hatted variables are those in the Einstein frame defined in Appendix A
and we have used $\hatH\simeq H/\sqrt{\Om}$ and $\hath^{ab}\simeq \Om
h^{ab}$ in the slow-roll approximation.  $N(\phi)$ is the e-folding
number defined by
\beqa
N(\phi)=\int_{t(\phi)}^{t_e}Hdt,
\eeqa
where $t_e$ is the time at the end of inflation. $N$ is conformally
invariant in the slow-roll approximation. The following useful relation
is immediately obtained
\beqa
H=-N_{,a}\dot\phi^a\simeq N_{,a}\frac{V_{eff}^{,a}}{3H},
\label{ndot}
\eeqa
where the slow-roll equation Eq.(\ref{eq:slow:2}) is used in the last
equality. Eq.(\ref{ndot}) can be formally solved in terms of $N_{,a}$ as
\beqa
N_{,a}=\frac{\ka^2 V V_{eff,a}}{\Om V_{eff,c}V_{eff}^{,c}}+\perp_a,
\label{ndot2}
\eeqa
where $\perp_a$ is a term orthogonal to $V_{eff,a}$.  Since $d\ln k=d\ln
aH\simeq Hdt$, to the first order in the slow-roll parameters, the
spectral index of scalar perturbation is then given by
\beqa
n_S-1\equiv\frac{d\ln P_{\R}}{d\ln k}=2\frac{\dot H}{H^2}+2\frac{N^{,a}\dot N_{,a}}{HN^{,c}N_{,c}}.
\label{ns:1}
\eeqa
The first term in the right-hand-side of Eq.(\ref{ns:1}) is given by
Eq.(\ref{hdot}) and the second term can be simplified using the
following relations \cite{ns}
\beqa
\dot N_{,a}&=&\dot\phi^b\nabla_b\nabla_aN
=\nabla_a(\dot\phi^b\nabla_b N)-(\nabla_a\dot\phi^b)(\nabla_b N)
=-H_{,a}-N^{,b}\nabla_a\dot\phi_b, \\
\nabla_a\dot\phi_b&\simeq& -\nabla_a(V_{eff,b}/3H)
=\frac{H_{,a}V_{eff,b}}{3H^2}-\frac{\nabla_a\nabla_b V_{eff}}{3H}.
\eeqa
Hence Eq.(\ref{ns:1}) can be written as\footnote{Unlike \cite{ss,ns},
the term proportional to the Riemann tensor of $h_{ab}$, which is higher
order in slow-roll approximations, does not appear here because the
slow-roll equation of motion is used to calculate $\nabla_a\dot\phi_b$.}
\beqa
n_S-1=-2\epsilon-\delta -\frac{2\ka^2}{\Om N_{,c}N^{,c}}-\frac{2\Om_{,a}N^{,a}}{\Om N_{,c}N^{,c}}
+\frac{2\eta_{ab}N^{,a}N^{,b}}{N_{,c}N^{,c}}. 
\label{index:scalar}
\eeqa
%
The expression can be further simplified using Eq.(\ref{ndot2}) for the
case when $N={\rm constant}$ hypersurfaces coincide with $V_{eff}={\rm
constant}$ hypersurfaces \cite{yst}.  In this case $\perp_a$ in
Eq.(\ref{ndot2}) is vanishing, which corresponds to neglecting the
isocurvature mode \cite{gwbm}\footnote{{Note that these modes are
initially isocurvature, but they could be adiabatic later like a
curvaton and/or a modulated reheating scenario.}}, and we obtain a very
simple formula for the scalar spectral index
\beqa
n_S-1=-6\epsilon-3\delta+2\eta_{ab}M^{ab},\label{index:scalar:simplified}\\
M^{ab}=\frac{N^{,a}N^{,b}}{N^{,c}N_{,c}}=
\frac{V_{eff}^{,a}V_{eff}^{,b}}{V_{eff,c}V_{eff}^{,c}}={1\over \Om}\widehat{M}^{ab},
\eeqa
which looks quite similar to that for a single field \cite{cy}.
Moreover, using the relation Eqs.(\ref{slowroll:e1}) and
(\ref{slowroll:e2}), one may confirm the conformal invariance of the
spectral index
\beqa
n_S-1=-6\epsilon-3\delta+2\eta_{ab}M^{ab}=-6\hatep+2\hateta_{ab}\widehat{M}^{ab}=\widehat{n_S}-1.
\label{index:scalar:inv}
\eeqa

Tensor perturbations are not changed due to the extension to multiple
fields.  The tensor power spectrum $P_{h}(k)$ is given by
\beqa
P_h(k)={8\ka^2}\left(\frac{\hatH}{2\pi}\right)^2=\frac{8\ka^2}{\Om}\left(\frac{H}{2\pi}\right)^2.
\label{spec:tensor}
\eeqa
Then the tensor spectral index is given by
\beqa
n_T\equiv \frac{d\ln P_{h}}{d\ln k}=-2\epsilon=-2\hatep=\widehat{n_T},
\label{index:tensor}
\eeqa
which is of course conformally invariant.  The tensor to scalar ratio
$r$ is also calculated as
\beqa
r\equiv\frac{P_h}{P_{\R}}=\frac{8\ka^2}{\Om N_{,c}N^{,c}}=
\frac{8\ka^2}{\hath^{ab}\hatN_{,a}\hatN_{,b}}=\widehat{r},
\label{tensor-scalar}
\eeqa
which is again conformally invariant.  $r$ can be written using Eq.(\ref{ndot2}) as
\beqa
r=\frac{16\epsilon}{1+\frac{2\Om\epsilon}{\ka^2}\perp_c\perp^c}\leq 16\epsilon= -8n_T,
\eeqa
where we have assumed the positivity of the scalar field metric $h_{ab}$ so that 
$\perp_c\perp^c\geq 0$. 
For the case when $N={\rm constant}$ hypersurfaces coincide with
$V_{eff}={\rm constant}$ hypersurfaces, we have $r=-8n_T$.  However in
general due to the presence of isocurvature mode corresponding to
$\perp_a$ in Eq.(\ref{ndot2}) the equality becomes an inequality: $r\leq
-8n_T$. These results are summarized in Table 1.

Here, it should be noted that $V_{eff}={\rm constant}$ hypersurfaces
coincide with $\widehat{V} ={\rm constant}$ hypersurfaces. Therefore,
the decomposition into adiabatic and isocurvature modes both in the
Jordan frame and the Einstein frame also coincide, which, together with
the conformal invariance of the curvature perturbation $\zeta$
\cite{cy,ms}, leads to the conclusion that we cannot discriminate the
frames by the observations of the primordial fluctuations up to the
leading order.

\begin{table}
  \begin{center}
  \setlength{\tabcolsep}{3pt}
  \begin{tabular}{|c|c| c|} 
   &Jordan frame $g_{\mu\nu}$ & Einstein frame $\hatg_{\mu\nu}=\Omega g_{\mu\nu}$
\\ \hline
 slow-roll parameters & $\epsilon,\eta_{ab},\delta$ & 
$\hatep=\epsilon,\hateta_{ab}$
 \\ 
 scalar spectral index  $n_S$& $1-6\epsilon-3\delta+2\eta_{ab}M^{ab}$& $1-6\hatep+2\hateta_{ab}M^{ab}$\\
tensor spectral index $n_T$& $ -2\epsilon$& $-2\hatep$\\
tensor/scalar ratio $r$ & $16\epsilon=-8n_T$& $16\hatep=-8\widehat{n_T}$\\    
  \end{tabular}
  \end{center}
\caption{Slow-roll parameters and inflationary observables in Jordan/Einstein frame 
for the case when $N={\rm constant}$ hypersurfaces coincide with $V_{eff}={\rm constant}$ 
hypersurfaces.}
\label{tab1}
\end{table}

\subsection{Example: Extended Assisted Inflation}

As an example of multi-scalar extended slow-roll inflation, we consider the 
assisted inflation model \cite{lms,ko}: $n$ scalar fields $\p^a$ each with an 
identical potential $V(\p^a)$. We assume a flat scalar field metric, $h_{ab}=\delta_{ab}$. 

For minimally coupled scalar fields, it was shown that inflation can
proceed even if each of the individual fields has a potential too steep
to sustain inflation on its own \cite{lms,ko}.  This can be immediately
seen from the action Eq.~(\ref{action}).  First, consider a set of $n$
minimally coupled $(\Om=1)$ scalar fields each with the same potential
\beqa 
S=\int d^4x\sqrt{-g}\left[{1\over 2\ka^2}R-
\sum_{a=1}^{n}\left({1\over 2}(\del\phi^a)^2+V(\phi^a)\right)\right].
\eeqa 
Since the cross coupling between different fields is absent, the
system is equivalent to $n$ copies of the same field and the action is
rewritten as 
\beqa 
S&=&\int d^4x\sqrt{-g}\left[{1\over 2\ka^2}R- 
{n\over2}(\del\phi^1)^2-nV(\phi^1)\right] \nonumber\\ 
&=&\int d^4x\sqrt{-g}\left[{1\over 2\ka^2}R- {1\over
2}(\del\widetilde{\phi}^1)^2-\widetilde{V}(\widetilde{\phi}^1)\right], 
\eeqa 
where
\beqa 
\widetilde{\p}^1=\sqrt{n}\p^1,~~~\widetilde{V}=nV.  \label{redefinition}
\eeqa 
Thus, the system is equivalent to a single scalar field with a potential
of the same form.  For example, for an exponential potential,
$V(\p)=V_0\exp(-\lambda \ka\p)$, the redefined potential becomes,
\beqa
\widetilde{V}(\widetilde{\p})= nV_0\exp\left(-(\lambda/\sqrt{n})
\ka\widetilde{\p}\right)=:\widetilde{V_0}\exp(-\widetilde{\lambda}\ka\widetilde{\p}).
\label{reduced:1}
\eeqa 
Hence for large $n$ the slope of the
potential $\widetilde{V}(\widetilde{\p})$ becomes less steep than that
of $V(\p)$ \cite{lms}. {In fact, the slow-roll parameters are given
by $\widetilde{\epsilon} = \widetilde{\lambda}^2/2=\lambda^2/(2n) = \epsilon_1/n$ and
$\widetilde{\eta} = \widetilde{\lambda}^2=\lambda^2/n = \eta_1/n$, where $\epsilon_1$ and $\eta_1$ are 
the slow-roll parameters of a single field, and hence are suppressed by
$n$. Then, the observable quantities are given by ${n}_{S}-1 =
{n}_{T} = -r/8 = -\lambda^2/n.$} For a
monomial potential, $V(\p)=\lambda \p^p$, the redefined potential becomes, 
\beqa
\widetilde{V}(\widetilde{\p})=(\lambda/n^{p/2-1})\widetilde{\p}^p
=:\widetilde{\lambda}\widetilde{\p}^p,
\label{reduced:2}
\eeqa
and for large $n$ the self-interactions become weaker
for $p>2$ \cite{ko}. {In this case, however, the slow-roll parameters
are not changed and the observable quantities are given by
${n}_{S}-1 =-p(p+2)/(\ka^2\widetilde{\p}^2)= -(p+2)/(2N)$ and ${n}_{T} =
-{{r}}/8 =-p^2/(\ka^2\widetilde{\p}^2)= -p/(2N)$, with $N$ being the e-folding number until 
the end of inflation.}

Even if a non-minimal coupling is introduced, assisted inflation can
persist and a redefinition Eq.~(\ref{redefinition}) again helps to flatten the potential but
does not help to weaken the non-minimally coupling.  {To be
specific, consider the $n$ copies of non-minimally coupled with a
potential with $\Om(\p)=1-\xi\ka^2\sum_{a=1}^n(\p^a)^2={1\over
n}\sum_{a=1}^n\widetilde{\Om}(\p^a)$ with
$\widetilde{\Om}(\p^a)=1-n\xi\ka^2(\p^a)^2$
\beqa
S&=&\int d^4x\sqrt{-g}\left[{\Om(\p)\over 2\ka^2}R-
\sum_{a=1}^n\left(
{1\over 2}(\del\phi^a)^2+V(\phi^a)\right)\right] \nonumber\\
&=&\int d^4x\sqrt{-g}\left[\sum_{a=1}^n\left({\widetilde{\Om}(\p^a)\over 2n\ka^2}R-
{1\over 2}(\del\phi^a)^2-V(\phi^a)\right)\right] \nonumber\\
&=&\int d^4x\sqrt{-g}\left[{\widetilde{\Om}(\p^1)\over 2\ka^2}R-
{n\over 2}(\del\phi^1)^2-nV(\phi^1)\right].
\eeqa
Thus, the system is equivalent to a non-minimally coupled single field
with the conformal factor and the potential of the same form.  Similar
to the minimally coupled case, the field redefinition
Eq.~(\ref{redefinition}) does help to flatten the potential as in Eq.~(\ref{reduced:1}) 
and Eq.~(\ref{reduced:2}), but the non-minimal coupling parameters $\xi$ is the same:
$\widetilde{\Om}(\widetilde{\p})=1-n\xi\ka^2\p^2=1-\xi\ka^2\widetilde{\p}^2$.} 
For example, for $V(\p)=\lambda \p^p$, observational quantities become \cite{cy} 
$n_S-1=n_T=-r/8\simeq (p-4)^2\xi$, for $|\xi|\ka^2\widetilde{\p}^2\gg 1$ and $\xi<0$, if $|\xi|\ll 1$ 
and $p\neq 4$. Because of the invariance of $\xi$, the slow-roll parameters and $n_S,n_T$ and $r$ 
are not changed in this case.

\section{Slow-Roll Inflation with Generalized Gravity}

The action we consider is \cite{Hwang:1996np}
\beq
S=\int d^4x\sqrt{-g}\left[{f(\phi,R)\over 2\ka^2}-
{1\over 2}\omega(\phi)(\nabla \phi)^2-V(\phi)\right].
\label{action2}
\eeq
Firstly, we show that the action is equivalent to that of two scalar
fields non-minimally coupled to gravity with degenerate scalar field
metric $h_{ab}$.  By introducing auxiliary field $\psi$,
Eq.(\ref{action2}) is dynamically equivalent to \cite{chiba}
\beqa
S=\int d^4x\sqrt{-g}\left[{f(\phi,\psi)\over 2\ka^2}+\frac{f_{,\psi}}{2\ka^2}(R-\psi)-
{1\over 2}\omega(\phi)(\nabla \phi)^2-V(\phi)\right],
\label{action3}
\eeqa
where $f_{,\psi}=\del f/\del \psi$. One can easily verify that the
equation of motion for $\psi$ gives $\psi=R$ if $f_{,\psi\psi}\neq 0$,
which reproduces the original action.  Eq.(\ref{action3}) is the action
of two scalar fields non-minimally coupled to gravity ($\Om=f_{,\psi}$)
with the unusual metric $h_{ab}$: $h_{\phi\phi}=\omega, h_{\psi\psi}=0$.
Thus, the generalized gravity theory is equivalent to two scalar fields
theory non-minimally coupled to gravity \cite{maeda,fm2}.

Assuming that the universe is described by the flat, homogeneous, and
isotropic universe model with the scale factor $a$, the field equations
derived from Eq.(\ref{action3}) are  given by
\beqa
H^2\Om+H\dot\Om=\frac{\ka^2}{3}\left[\frac{1}{2}\omega\dot\p^2+V
+\frac{1}{2\ka^2}\left(\psi\Om-f\right) \right]=:
\frac{\ka^2}{3}\left(\frac{1}{2}\omega\dot\p^2+U\right)
,\label{eq:3}\\
\ddot\Om-H\dot\Om+2\dot H\Om=-\ka^2\omega\dot\p^2,\label{eq:4}\\
\ddot\p+3H\dot\p=-\frac{1}{\omega}
 \left({1\over 2}\omega_{,\p}\dot{\p}^2+V_{,\p}-\frac{f_{,\p}}{2\ka^2}\right)
=-\frac{1}{\omega}
 \left({1\over 2}\omega_{,\p}\dot{\p}^2+U_{,\p}-\frac{\psi\Om_{,\p}}{2\ka^2}\right),
\label{eq:5}
\eeqa
where $\Omega=f_{,\psi}$ is the conformal factor. 

\subsection{Slow-Roll Conditions}

As done in the previous section, we assume that the time scale of the
motion of the scalar field is much larger than the cosmic time scale
$H^{-1}$ under the slow-roll approximations. Then, as slow-roll
conditions of the scalar field in the generalized gravity, we impose
that $\omega \dot\p^2 \ll |U|, |\dot\Om|\ll H\Om$,
$|\ddot\p+\omega_{\p}\dot{\p}^2/(2\omega)|\ll H|\dot\p|$,
$|\ddot\p+\omega_{\p}\dot{\p}^2/(2\omega)| \ll |U_{,\p}/\omega|$, which
yields
\beqa
&&H^2\Om\simeq \frac{\ka^2}{3}U,
\label{eq:geneslow:1}\\
&&3H \dot\p\simeq -\frac{U_{,\p}}{\omega}+6\frac{\Om_{,\p}H^2}{\ka^2\omega}
\simeq -\frac{\Om^2}{\omega}\left(\frac{U}{\Om^2}\right)_{,\p}
=:-V^{eff}_{,\p},
\label{eq:geneslow:2}
\eeqa
where in the second equation, we have assumed $|\dot H/H^2|\ll 1$, which
is necessary to cause inflation and should be checked later. This
amounts to assuming that only $\p$ becomes an inflaton and $\psi$ is a
dependent variable of $\p$.\footnote{To be more detailed, we may
regard Eq.~(\ref{eq:geneslow:1}) as the equation of the variables $\p$
and $\psi$, that is, $\psi \simeq 4 \ka^2 U(\p,\psi)$ and solve it
implicitly to express $\psi= \psi(\p)$ \cite{Yamaguchi:2005qm}.  Then,
by inserting this expression $\psi$ into the original potential
$U(\p,\psi)$ and the conformal factor $\Om(\p,\psi)$, we obtain the
reduced potential $U(\p,\psi(\p))$ and the reduced conformal factor
$\Om(\p,\psi(\p))$.}

In the following, we derive the consistency conditions for the
slow-rolling of the scalar field in the generalized gravity (the
generalized slow-roll conditions). By computing
$\ddot\p+\omega_{\p}\dot{\p}^2/(2\omega)$ from
Eq.~(\ref{eq:geneslow:2}), we obtain
\beqa
\frac{\ddot\p+\omega_{\p}\dot{\p}^2/(2\omega)}{H\dot\p}\simeq 
 -\frac{\dot H}{H^2} \left(1+\frac{8\ka^2UV^{eff}_{,\p\psi}}{\Om V^{eff}_{,\p}}\right)
 -\frac{\Om}{\kappa^2 U} \left(V^{eff}_{,\p\p}+\frac{\omega_{,\p}}{2\omega}V^{eff}_{,\p} \right),
\label{generatio:1}
\eeqa
where we have used $\dot\psi=\dot R\simeq 24H\dot H$. 
Also, from Eq.~(\ref{eq:geneslow:2}), we note
\beqa
&& \frac{\ddot\p+\omega_{\p}\dot{\p}^2/(2\omega)}{U_{,\p}}
  \simeq -\frac{\ddot\p+\omega_{\p}\dot{\p}^2/(2\omega)}{3H\dot\p}
         \frac{V^{eff}_{,\p}}{U_{,\p}}.
\eeqa
Moreover, from Eqs.~(\ref{eq:geneslow:1}) and (\ref{eq:geneslow:2}),
\beqa
&&\left| \frac{\omega \dot\p^2}{U} \right| \simeq 
  \frac{\omega\Om (V^{eff}_{,\p})^2}{3\ka^2 U^2},\label{generatio:2}\\
&&\frac{\dot\Om}{H\Om}\simeq -\frac{\Om_{,\p}V^{eff}_{,\p}}{\ka^2 U}
  +\frac{2\Om_{,\psi}\psi}{\Om}\frac{\dot{H}}{H^2}.\label{generatio:3}
\eeqa  
Hence, we introduce the following three slow-roll parameters and obtain
the generalized slow-roll conditions:
\beqa
&&\epsilon^{g}:=\frac{\omega\Om (V^{eff}_{,\p})^2}{2\ka^2 U^2};
~~~~\epsilon^{g} \ll 1, \label{geneslow-roll:1}\\
&&\eta^{g}:=\frac{\Om}{\ka^2 U}\left(V^{eff}_{,\p\p}+\frac{\omega_{,\p}}{2\omega}V^{eff}_{,\p}\right);~~~~|\eta^{g}|\ll1,
\label{geneslow-roll:2}\\
&&\delta^{g}:=\frac{\Om_{,\p}V^{eff}_{,\p}}{\ka^2 U};~~~~|\delta^{g}|\ll1,
\label{geneslow-roll:3}
\eeqa
and three subsidiary conditions:
\beqa
\left| \frac{\kappa^2UV^{eff}_{,\p\psi}}{\Om V^{eff}_{,\p}} \right|= {\cal O}(1),
\label{genesub:1}\\
\left| \frac{\kappa^2\Om_{,\psi}U}{\Om^2}\right|= {\cal O}(1),\label{genesub:2}\\
\left| \frac{U_{,\p}}{\omega V^{eff}_{,\p}}\right|= {\cal O}(1).\label{genesub:3}
\eeqa
Note that since from Eq.~(\ref{eq:geneslow:1}) $\dot H/H^2$ is approximated as
\beqa
 \frac{\dot H}{H^2} \simeq -\frac{\epsilon^{g}+{1\over 2}\delta^{g}}
{1-\frac{\Om_{,\psi}\psi}{\Om}}.
\label{genehdot}
\eeqa
Thus, $|\dot H/H^2|\ll 1$ is guaranteed by these slow-roll conditions
unless $1-\Om_{,\psi}\psi/\Om$ becomes vanishing accidently.\footnote{It
is interesting to note that $\Om_{,\psi}\psi=\Om$ implies
$f(\p,R)=f(\p)R^2+g(\p)$ with $f(\p)$ and $g(\p)$ being arbitrary
functions of $\p$, which corresponds to $R^2$ inflation model
\cite{fm2}.  In this case, both $\p$ and $\psi$ become inflatons and
higher order corrections of slow-roll approximations become
important. {However, in the case when $f(\p,\psi)=f(\psi)$,
only $\psi$ can be an inflaton and the primordial fluctuations can be
calculated in the Einstein frame as given in 
\ref{subsub:fR}.}}

To sum up, the generalized slow-roll conditions consist of three
conditions Eqs.~(\ref{geneslow-roll:1}), (\ref{geneslow-roll:2}),
(\ref{geneslow-roll:3}) and three subsidiary condition
Eqs.~(\ref{genesub:1}), (\ref{genesub:2}), and
(\ref{genesub:3}).\footnote{It should be noted that three subsidiary
conditions are the sufficient conditions for slow-roll and the necessary
conditions are that Eq.~(\ref{genesub:1}) (or Eq.~(\ref{genesub:2}) or
Eq.~(\ref{genesub:3})) multiplied by $\dot{H}/H^2$ are sufficiently
small.}

As an example, we consider the case of a non-minimal coupling $f = R
\Omega(\p)$. In this case, $U(\p,\psi)= V(\phi)$ and $V^{eff}(\p,\psi) =
V^{eff}(\p)$. Then, all slow-roll conditions coincide with those in the
previous section.

\subsection{Perturbations}

In this subsection, we would like to evaluate the primordial density
fluctuations generated in the generalized gravity theory.  Since only
$\p$ is an independent field to the first order in the slow-roll
approximations, the system is equivalent to one scalar field ($\p$)
non-minimally coupled to gravity,    which has been already studied \cite{cy}.

From the conformal invariance of the curvature perturbation $\R$, the
power spectrum $P_{\R}(k)$ in the Jordan frame is given by that in the
Einstein frame \cite{alex,ss,ns,ms},
\beqa
P_{\R}(k)=\left(\frac{\hatH}{2\pi}\right)^2
\hatN_{,\hatp}^2
=\left(\frac{H}{2\pi}\right)^2\frac{1}{\omega}
N_{,\p}^2,
\label{genespec:scalar}
\eeqa
where $k$ is a comoving wavenumber at the horizon exit ($k=aH$) and the
hatted variables are those in the Einstein frame defined in Appendix A and we have used
$d\hatp =d\p\sqrt{\omega/\Om}$ and $\hatH\simeq H/\sqrt{\Om}$ in the
slow-roll approximation.  $N(\phi)$ is conformally invariant in the
slow-roll approximation, $\hatN = N$. Then, using the relation
Eqs.(\ref{geneslowroll:e1}) and (\ref{geneslowroll:e2}), the spectral
index of scalar perturbation is given by
\beqa
  n_S - 1 = -6\epsilon^{g}+2\eta^{g}-3\delta^{g}
          = -6\hatep^{g}+2\hateta^{g}
          = \widehat{n}_S-1,
\eeqa
which proves the conformal invariance of the spectral index
\cite{cy}. The tensor spectral index $n_T$ and the tensor to scalar
ratio $r$ are also conformal invariant \cite{cy}.

\section{Summary}

By extending our previous study, we have derived the slow-roll
conditions for multiple scalar fields non-minimally coupled to
gravity. The slow-roll conditions consist of three main conditions
Eqs.~(\ref{slow-roll:1}), (\ref{slow-roll:2}), (\ref{slow-roll:3}) and
one subsidiary condition Eq.~(\ref{sub:1}). We
have given the simple formulae of spectral indices and the tensor-to
scalar ratio, Eq.~(\ref{index:scalar}), Eq.~(\ref{index:tensor}) and
Eq.~(\ref{tensor-scalar}).  The scalar spectral index can be further
simplified and be written in terms of the slow-roll parameters if the
isocurvature perturbation is negligible,
Eq.~(\ref{index:scalar:simplified}).

We have also derived the slow-roll conditions for generalized gravity
theories and found that, to the first order in the slow-roll
approximations, the system is reduced to a single field with non-minimal
coupling. The formulae of the spectral indices are thus the same as our
previous results \cite{cy}.

We hope that our formulae may be useful to connect the inner space to
the outer space.

\ack
This work was supported in part by Grant-in-Aid for Scientific Research
from JSPS (No.\,17204018 (T.C.), No.\,20540280 (T.C.), No.\,18740157 (M.Y.), and No.\,19340054 (M.Y.))
and from MEXT (No.\,20040006(T.C.)) and in part by Nihon University.

\appendix
\section{Slow-Roll Conditions and Perturbations in Einstein Frame}
\label{app1}

In this appendix, we perform the conformal transformation to the
Einstein frame and introduce slow-roll parameters and give their
relations with those in the Jordan frame.

\subsection{A Non-minimally Coupled Multi-Scalar Field}

First, we consider the case of non-minimally coupled multiple
scalar fields. Introducing the Einstein metric $\hatg_{\mu\nu}$ by the
conformal transformation
\beqa
\hatg_{\mu\nu}=\Om(\p)g_{\mu\nu},
\label{conf}
\eeqa
the action Eq.(\ref{action}) becomes that of two scalar fields minimally
coupled to Einstein gravity
\beqa
S&=&\int d^4x\sqrt{-\hatg}\left[{1\over 2\ka^2}\widehat{R}-
\frac{1}{2\Om} \lmk h_{ab}+\frac{3\Om_{,a}\Om_{,b}}{2\ka^2\Om} \rmk
\hatg^{\mu\nu}\del_{\mu}\p^a \del_{\nu}\p^b
-\frac{V}{\Om^2}\right]\\
&=&\int d^4x\sqrt{-\hatg}\left[{1\over 2\ka^2}\widehat{R}-
{1\over 2}\widehat{h}_{ab}\hatg^{\mu\nu}\del_{\mu}\p^a \del_{\nu}\p^b
-\hatV(\p)\right],
\eeqa
where $\hath_{ab}$ and $\hatV$ are defined by
\beqa
\hatV(\p)=\frac{V(\p)}{\Om(\p)^2}\label{corres:2},\\
\hath_{ab}={1\over \Om}\left(h_{ab}+\frac{3\Om_{,a}\Om_{,b}}{2\ka^2\Om}\right).
\eeqa
%
{Here, it should be noted that one cannot define $\hatp^{a}$ by 
$\left(d\hatp^{a}\right)^2 \equiv \left(d\p^{a}\right)^2/\Om(\phi^a)$
because $dd\hatp^{a} \ne 0$, different from a single field case.}

\subsubsection{Slow-Roll Conditions}

The slow-roll conditions in the Einstein frame are simply given by
\beqa
&&\hatep:=\frac{1}{2\ka^2\hatV^2}\hath^{ab}\hatV_{,a}\hatV_{,b} ;~~~\hatep\ll 1,\label{eq:slow-ein}\\
&&\hateta_{ab}:=  
\frac{\widehat{\nabla}_a\widehat{\nabla}_b\hatV}{\ka^2\hatV} ;~~~~~~~~~~|{{\hateta}_{a}}^{b}|\ll 1.\\
\label{eq:slow-ein2}
\eeqa
Using Eq.~(\ref{corres:2}) and Eq.~(\ref{eq:slow:2}), these parameters
are rewritten as
\beqa
&&\hatep=\frac{\hath^{ab}V_{eff,a}V_{eff,b}}{2\ka^2 V^2},\\
&&\hateta_{ab}=\frac{\Om^2}{\ka^2V}\hatnabla_a\left(\frac{\nabla_bV_{eff}}{\Om^{2}}\right),
\eeqa
where $V_{eff}^{,a}$ is defined in Eq.~(\ref{eq:slow:2}).  If we assign
${\cal O}(\varepsilon/\sqrt{n})$ to $\sqrt{\Om}V_{eff,a}/\ka V$ and to
$\Om_{,a}/\ka \sqrt{\Om}$, then from Eq.(\ref{slow-roll:1}) and
Eq.(\ref{slow-roll:3}), we have $\epsilon={\cal O}(\varepsilon^2)$ and
$\delta={\cal O}(\varepsilon^2)$. Therefore, under the slow-roll
conditions in the Jordan frame, $\Om_{,a}\Om_{,b}/\ka^2\Om={\cal
O}(\varepsilon^2/n)$ and $\hath_{ab}\simeq h_{ab}/\Om$ is implied. Hence
the slow-roll parameters in the Einstein frame are related to the
slow-roll parameters in the Jordan frame as
\beqa
&&\hatep\simeq \epsilon,\label{slowroll:e1}\\
&&\hateta_{ab}\simeq {1\over \Om}\left(\eta_{ab}- \frac12 h_{ab}\delta-
\frac{3\Om_{,a}V_{eff,b}}{2\ka^2 V}+\frac{\Om_{,b}V_{eff,a}}{2\ka^2 V}
\right),
\label{slowroll:e2}
\eeqa
where we have used the relation \cite{wald}
\beqa
&&\hatnabla_av_b=\nabla_av_b-C^c_{ab}v_c, \\
&&C^c_{ab}=\frac12\left(h_{ab}\nabla^c\ln \Om-\delta^c_a\nabla_b\ln\Om-
\delta^c_b\nabla_a\ln\Om  \right).
\eeqa

\subsection{Generalized Gravity}

Next, we consider the case of the generalized gravity with a single
scalar field Eq.(\ref{action2}). From the equivalent action
Eq.(\ref{action3}), introducing the Einstein metric $\hatg_{\mu\nu}$ by
the conformal transformation
\beqa
\hatg_{\mu\nu}=f_{,\psi}(\p,\psi)g_{\mu\nu}\equiv\Om(\p,\psi)g_{\mu\nu},
\label{conf2}
\eeqa
the equivalent action Eq.(\ref{action3}) becomes that of two scalar fields minimally
coupled to Einstein gravity
\beqa
S=\int d^4x\sqrt{-\hatg}\left[{1\over 2\ka^2}\widehat{R}
-\frac{\omega}{2\Om}(\hatnabla\p)^2
-\frac{3}{4\ka^2\Om^2}(\hatnabla\Om)^2
-\frac{V}{\Om^2}-\frac{\psi\Om-f}{2\ka^2\Om^2}\right].
\eeqa

\subsubsection{Slow-Roll Conditions}

Only $\p$ can be an inflaton (see Eq.~(\ref{eq:geneslow:2})) and
$\psi\simeq 12H^2$ is a dependent variable of $\p$, to the first order
in slow-roll approximations.\footnote{Note that from the slow-roll equation of motion, 
$\psi\simeq 12H^2\simeq 4\ka^2U/\Om$, $\hatV_{,\psi}\simeq 0$ under the slow-roll approximation. 
So $\psi$ stays the minimum if it is massive, being consistent with our assumption. 
We do not consider the case when $\psi$ is light so that both $\p$ and $\psi$ can be inflatons. 
For the case when only $\psi$ can be an inflaton, see below.}  
 In the case when $\psi=\psi(\p)$, the
action becomes
\beqa
S&=&\int d^4x\sqrt{-\hatg} \left[{1\over 2\ka^2}\widehat{R}
-\frac{\omega}{2\Om}\left(1+\frac{3\Om_{,\p}^2}{2\ka^2\omega\Om}\right)(\hatnabla\p)^2
-\frac{U}{\Om^2}\right] \nonumber\\
&=&\int d^4x\sqrt{-\hatg}
\left[{1\over 2\ka^2}\widehat{R}
-\frac{1}{2}(\hatnabla\hatp)^2
-\hatV(\hatp)\right],
\eeqa
where $U$ is defined by Eq.~(\ref{eq:3}) and we have introduced a
canonically normalized scalar field $\hatp$ with a potential
$\hatV(\hatp)$ 
\beqa
&&(d\hatp)^2= \frac{\omega}{\Om}\left(1+\frac{3\Om_{,\p}^2}{2\ka^2\omega\Om}\right)(d\p)^2,\\
&&\hatV=\frac{U}{\Om^2}=
\frac{1}{\Om^2}\left(V+\frac{\psi\Om-f}{2\ka^2}\right).
\label{hatV}
\eeqa
The slow-roll parameters and conditions in the Einstein frame are simply
\beqa
&&\hatep_g:=\frac{(\hatV_{,\hatp})^2}{2\ka^2\hatV^2};~~~~~\hatep_g\ll 1,\\
&&\hateta_g:=\frac{\hatV_{,\hatp\hatp}}{\ka^2\hatV};~~~~~~~|\hateta_g|\ll 1.
\eeqa
If we assign ${\cal O}(\varepsilon^2)$ to the slow-roll parameters
($\epsilon_g,\eta_g,\delta_g$), then
$3\Om_{,\p}^2/(2\ka^2\omega\Om)={\cal O}(\varepsilon^2)$, and
$d\hatp^2\simeq \omega d\p^2/\Om$ is satisfied under the slow-roll
conditions Eq.(\ref{geneslow-roll:1}) and Eq.(\ref{geneslow-roll:3}).
Therefore, under the slow-roll approximations, the slow-roll parameters
in the Einstein frame are related to those in the Jordan frame as
\cite{cy}
\beqa
&&\hatep_g\simeq \epsilon_g,\label{geneslowroll:e1}\\
&&\hateta_g\simeq \eta-\frac{3}{2}\delta.\label{geneslowroll:e2}
\eeqa

\subsubsection{$f(R)$ Inflation}

\label{subsub:fR}

 Different from the previous subsection, we consider the case that
only $\psi$ exists and is an inflaton, that is,
$f(\p,\psi)=f(\psi)$ and $\omega(\p)=V(\p)=0$. 
Then, the equivalent action Eq.(\ref{action3}) in the Einstein frame becomes
\beqa
S&=&\int d^4x\sqrt{-\hatg}\left[{1\over 2\ka^2}\widehat{R}
-\frac{3\Om_{,\psi}^2}{4\ka^2\Om^2}(\hatnabla\psi)^2
-\frac{\psi\Om-f}{2\ka^2\Om^2}\right] \nonumber \\
&=&\int d^4x\sqrt{-\hatg}\left[{1\over 2\ka^2}\widehat{R}
-\frac12 (\hatnabla\hatpsi)^2
-\hatV(\hatpsi)\right],
\eeqa
where $(d\hatpsi)^2 = 3\Om_{,\psi}^2(d\psi)^2/(2\ka^2\Om^2)$ and
$\hatV(\hatpsi) = (\psi\Om-f)/(2\ka^2\Om^2)$. Then, the slow-roll parameters and conditions 
are given by
\beqa
&&\hatep_{g}:=\frac{(\hatV_{,\hatpsi})^2}{2\ka^2\hatV^2};~~~~~\hatep_{g}\ll 1,\\
&&\hateta_{g}:=\frac{\hatV_{,\hatpsi\hatpsi}}{\ka^2\hatV};~~~~~~~|\hateta_{g}|\ll 1.
\eeqa
Here $\hatV_{,\hatpsi}$ and $\hatV_{,\hatpsi\hatpsi}$ are given by
\beqa
\hatV_{,\hatpsi}=\frac{1}{\sqrt{6}\ka\Om}\left(-\psi+\frac{2f}{\Om}\right),\\
\hatV_{,\hatpsi\hatpsi}=\frac13\left[
\frac{2}{\Om}\left(\psi-\frac{2f}{\Om}\right)
+\frac{1}{\Om_{,\psi}}\left(1-\frac{\psi\Om_{,\psi}}{\Om}\right) \right].
\eeqa
It is to be noted that the condition $\hatV_{,\hatpsi} = 0$ is equivalent
to $f \propto \psi^2$. Furthermore, one can easily check that
$\hatV_{,\hatpsi\hatpsi} = 0$ for $f \propto \psi^2$. Thus, as is well known,
only the models with $f(\psi)$ which slightly deviates from $\psi^2$ can
cause inflation. 

As examples, we consider two models. The first one is a power law model: 
$f(\psi) = A \psi^{2+\alpha}$ ($\alpha \ll 1, A$ : constants). Then, the
slow-roll parameters are given by
\beqa
&& \hatep_{g} = \frac{\alpha^2}{3(1+\alpha)^2},\\
&& \hateta_{g} = \frac{2\alpha^2}{3(1+\alpha)^2}.
\eeqa
Then, the observable quantities are given by
\beqa
&& \widehat{n}_{S}-1 =-6\hatep_{g}+2\hateta_{g} = -\frac{2\alpha^2}{3(1+\alpha)^2}, \\
&& \widehat{n}_{T} = -\frac{\widehat{r}}{8} =-2\hatep_{g} =
-\frac{2\alpha^2}{3(1+\alpha)^2}.
\eeqa
As the second model we consider the Starobinsky model \cite{alex80}: $f(\psi) = \psi+
\psi^2/(6M^2)$ ($M \ne 0$ : constant).  In terms of $\ka\hatpsi=\sqrt{\frac32}\ln \Om$, 
the potential $\hatV$ becomes
\beqa
\hatV=\frac{\psi^2}{12\ka^2M^2\left(1+{\psi\over 3M^2}\right)^2}=\frac{3M^2}{4\ka^2}
\left(1-e^{-\sqrt{\frac23}\ka\hatpsi}\right)^2.
\eeqa
The slow-roll parameters are then given by
\beqa
&& \hatep_{g} = \frac{12 M^4}{\psi^2} =\frac{4}{3\left(e^{\sqrt{\frac23}\ka\hatpsi}-1\right)^2}
\simeq \frac{3}{4N^2},\\
&& \hateta_{g} = \frac{4M^2(3M^2-\psi)}{\psi^2} =
\frac{4\left(2-e^{\sqrt{\frac23}\ka\hatpsi}\right)}
{3\left(e^{\sqrt{\frac23}\ka\hatpsi}-1 \right)^2}
\simeq -\frac{1}{N},
\eeqa
where the e-folding number $N$ is given by $N \simeq \psi/(4M^2)\simeq  
3e^{\sqrt{\frac23}\ka\hatpsi}/4$.
Then, the observable quantities are given by
\beqa
&& \widehat{n}_{S}-1 =-6\hatep_{g}+2\hateta_{g} \simeq -\frac{9}{2N^2}-\frac{2}{N}, \\
&& \widehat{n}_{T} = -\frac{\widehat{r}}{8} =-2\hatep_{g} =
-\frac{3}{2N^2},
\eeqa
in agreement with \cite{hn}. 

As a final remark,  it should be noted that using the slow-roll equation of
motion, $\psi\simeq 12H^2\simeq 12\hatH^2\Om\simeq 4\ka^2\Om\hatV$, we
find $\hatV_{,\hatpsi}\simeq 0$ up to the leading order of the slow-roll
approximations. That is, in order to relate these results in the
Einstein frame to those in the Jordan frame, one needs to go beyond the
leading order of slow-roll approximations.



\end{document}